# Energy delocalization in strongly disordered systems induced by the long-range many-body interaction

Alexander L. Burin

*Department of Chemistry, Tulane University, 6400 Freret Street, New Orleans, LA 70118-5636, USA*

**Anderson localization[1] in a random system is sensitive to a distance dependence of the excitation transfer amplitude *V(r)*. If *V(r)* decreases with the distance *r* slower than *1/r^d* in a *d*-dimensional system then all excitations are delocalized at arbitrarily strong disordering,[1,2] due to the resonant interaction of far separated quantum states (Fig. 1). At finite temperature *T>0* the density of excitations is finite and they can influence each other by means of their interaction. Many–body excitations involving simultaneous transitions of several single particle excitations create additional channels for energy delocalization and transport. Here we show that if the interaction of excitations decreases with the distance slower than *1/R^{2d}* then excitations are delocalized at finite temperature irrespectively to disordering.[3] This delocalization results in the finite decoherence rate in the ensemble of interacting spins *1/2* representing the model of quantum computer, thus restricting the quantum hardware performance. It also leads to the energy and particle delocalization and transport at finite temperature in various physical systems including doped semiconductors, despite of the full localization of electrons at zero temperature.**

Quantum particle is localized in a random potential if disordering is sufficiently large.[1] However, this conclusion fails if the transfer amplitude of excitation depends on the distance according to the power law

$$U(r) \approx \frac{U_0}{r^a}, \tag{1}$$

with the exponent $a$ less than the system dimension $d$.[1,2] In that case the coupling strength of quantum energy levels Eq. (1) decreases with increasing the system size $L$ as $L^{-a}$. At sufficiently large size $L$ it exceeds a characteristic splitting of quantum energy levels $\delta\varepsilon \approx W/L^d$ ($W$ is the disordering energy and $d$ is the system dimension). Therefore at sufficiently large system size $L$ each excited state establishes resonant couplings with far separated quantum states (see Fig. 1) and thus gets delocalized.

The above consideration is restricted to single particle excitations. The example of propagating single particle excitation is shown in Fig. 1. Initially excited particle (spin) $i$ goes to its ground state transferring the excess of energy to the particle $j$ which gets excited. When the density of excitations is finite (finite temperature) one also should consider the possibility of delocalization associated with many-body excitations in addition to single particle excitations. If there is more than one excited particle then simultaneous transitions involving two or more particles are possible. For instance, two particle excitations are formed by two simultaneous single-particle transitions (see Fig. 2). In the large system the number of many-body excitations exceeds that for the single particle excitations. For instance, each couple of excitations can form a two-particle excitation so the number of two particle excitations scales with the system size as $L^{2d}$ in contrast with $L^d$ for single particle excitations. The interlevel spacing for two particle excitations $\delta\varepsilon_2 \propto 1/L^{2d}$ is smaller than that for single particle excitations and therefore it can create additional opportunities for the energy delocalization. Below we demonstrate that at finite temperature the delocalization takes place inevitably in the system with the interaction Eq. (1) decreasing with the distance according to the power law $1/L^a$ with the exponent $a$ less than the doubled system dimension $d$ ($a<2d$).[3]

In this letter we study the very general model of interacting two-level excitations in *d*-dimensions in the regime of strong disordering such that all single-particle excitations are localized. All interactions obey the law Eq. (1) with the exponent $a \geq d$. The model of two-level excitations is relevant for various strongly disordered systems where each excited state can be strongly coupled with the only one state. For instance, in doped semiconductors with localized electrons two-level excitations are formed by closely located electron-hole pairs,[4] while far separated electron-hole pairs have negligibly small coupling because of the electron localization. Two-level model also describes the energy transport promoted by interacting two-level systems in amorphous solids[5] or magnetic molecules[6] at low temperature. This model describes quantum bits (qubits) in quantum computers. Qubits can be often coupled by the long-range interaction Eq. (1). Indeed they must be controllable by external electric or magnetic fields which requires them to possess electric or magnetic dipole moments both leading to their *1/r³* interaction.[7] The standard gate operation scenario of quantum computations[8] requires all involved qubits to stay in coherent superpositions of ground and excited states. Thus the number of excitations can be as large as the total number of qubits and many-body interaction of them can be significant. Delocalization of energy in the qubit ensemble destroys the quantum information and therefore it has to be avoided. Thus our study is significant for the design of quantum hardware.

Two level excitations can be modelled by spins *1/2*. Consider the system of interacting spins *1/2*. Each spin *i* is placed into a random external static field $\Delta_i$ along the local quantization axis *z* for each spin. We also assume that the interaction of pseudospins Eq. (1) contains longitudinal and transverse parts. The longitudinal interaction $U_{ij}$ couples z-projections of spins while the transverse interaction $V_{ij} \sim \xi U_{ij}$ couples *x*-projections of spins. The Hamiltonian of the system reads

$$\hat{H} = -\sum_i \Delta_i S_i^z - \frac{1}{2}\sum_{i \neq j} U_{ij} S_i^z S_j^z - \xi \sum_{i \neq j} V_{ij} S_i^x S_j^x, \ V_{ij} \sim U_{ij}, \ \xi < 1 \qquad (2)$$

This particular choice of the system Hamiltonian is not unique but it is quite general. Particularly, it accounts for many-body interactions.[5]

We assume that energies $\Delta_i$ are random and distributed uniformly in the domain $(-W, W)$ with the density $g=1/(2W)$. We also assume that pseudospins occupy a $d$-dimensional space and have a density $n$. Since the limit of strong disordering is considered we assume that the characteristic interaction of pseudospins $U_0 n^{a/d}$ is much smaller than the typical spin energy $\Delta_i \sim W$

$$U_0 n^{a/d} << W. \qquad (3)$$

Finally we assume the each spin has equal chances to be in the ground or excited states. This is the case of the infinite system temperature $T=\infty$ and it is most easy for study. At a finite temperature $T>0$ the system behavior is qualitatively similar and can be characterized considering only spins with the energy less than the thermal energy and setting for them approximately $T=\infty$.[5]

The transverse interaction leads to the energy transfer between two pseudospins by means of the flip-flop transition if one spin $i$ is initially in its excited state, while the other spin $j$ is in its ground state (Fig. 1). Since the interaction is weak Eq. (3) we can treat the transverse part of the Hamiltonian Eq. (2) as a perturbation. Then one can assume that all spins are approximately characterized by their $\pm 1/2$ projections onto the z-axis. Each transverse term $V_{ij}$ induces the flip-flop transition ($S_i^z = 1/2, S_j^z = -1/2 \rightarrow S_i^z = -1/2, S_j^z = 1/2$, see Fig. 1). Such transition results in the longitudinal energy change $\Delta_{ij} = \Delta_i - \Delta_j + \sum_k (U_{ik} - U_{jk}) S_k^z$. The flip-flop transition is efficient under the condition of resonance $V_{ij} \geq \Delta_{ij}$ where the transverse amplitude $V_{ij}$ compensates the change in the longitudinal energy $\Delta_{ij}$ as required by the energy conservation. Then the excitation is shared between sites $i$ and $j$ as follows from the wavefunctions of the flip-flop pair eigenstates

$$\psi_+ = \cos(\varphi_{ij}/2)|+-> + \sin(\varphi_{ij}/2)|-+>, \ |\pm\pm> = |S_i^z = \pm 1/2; S_i^z = \pm 1/2>$$
$$\psi_- = -\sin(\varphi_{ij}/2)|+-> + \cos(\varphi_{ij}/2)|-+>, \ \tan(\varphi_{ij}) = V_{ij}/(2\Delta_{ij}).$$
(4)

Since the longitudinal energy of the flip-flop transition $\Delta_{ij} \sim W$ is large compared to the interaction energy $V_{ij} \leq U_0 n^{a/d}$ (see Eq. (3)) the resonances are quite rare. One can characterize them by the probability $P_{ij}$ given by the ratio of the size of the resonant energy domain $(-V_{ij}, V_{ij})$ and the size of the overall energy domain $W$, which yields $P_{ij} = |V_{ij}|/W$. The delocalization of a single excitation can be characterized by the total number of resonances per the single excitation $N = \sum_j |V_{ij}|/W$. When the interaction Eq. (1) decreases with the distance slower then $1/r^d$ the number of resonances $N(L)$ in a d-dimensional system of the size $L$ tends to infinity with increasing the system size $L$ as follows from the estimate below

$$N(L) \sim n\int_l^L dr^d \frac{\xi U_0}{r^a W} \sim \frac{\xi U_0 n L^{d-a}}{W}.$$
(5)

Here $l \sim n^{-1/d}$ is the average distance between pseudospins and we do not consider the marginal case $a=d$.

The divergence of the number of resonances Eq. (5) leads to sharing of any local excitation $i$ with an infinite number of other spins $j$, which results in the delocalization of excitation energy.[1,2] One can describe delocalization transition as the formation of the infinite cluster of resonantly coupled sites. If such cluster is formed by sites separated from each other by some typical distance $L_*$, i. e. involving resonant sites separated from each other by the distance $R$, such as $L_* < R < \eta L_*$, and $\eta$ is the scaling parameter of order of $1$ then the number of spins coupled to the given spin $i$ must be of order of unity

$$\chi(L) \sim N(\eta L) - N(L) \sim \frac{dN}{d\ln(L)} = \chi_c \sim 0.1.$$
(6)

The effective delocalization parameter $\chi_c$ in Eq. (6) is of order *0.1* in accordance with the original Anderson's paper[1] and later numerical studies.[9,10] Irreversible flip-flop transitions takes place between adjacent spins belonging to the infinite cluster and the characteristic irreversible transition rate can be estimated using the transverse interaction at the average distance $L_*$ between adjacent resonances

$$\frac{1}{\tau} \sim \frac{\xi U_0}{\hbar L_*^a} \sim \frac{E_*}{\hbar}. \tag{7}$$

In our case of interest Eq. (3) when disordering is strong and $a \geq d$ a percolation cluster cannot be formed and single particle excitations are localized at any system size $L$.[2] However there are many such excitations at finite temperature and one should consider their interaction absent in the single-particle approximation. To investigate the most significant resonant interaction we study the coupling of resonant flip-flop pairs (see Eq. (4) and Fig. 1) which are most mobile single particle objects. These resonant pairs can be characterised by their size $L$ and energy $E$ given by their coupling energy $E=\xi U_0/L^a$. The density of such resonant pairs can be estimated as the product of the density of excited pseudospins $n$ and the probability of resonance at distance $\sim L$ which is given by the factor $\chi(L)$ Eq. (6)

$$n_p(L) = n\chi(L) \sim n \frac{\xi U_0 n L^{d-a}}{W} = n \frac{\xi U_0 n^{a/d}}{W} \left(\frac{E}{\xi U_0 n^{a/d}}\right)^{\frac{a-d}{a}}. \tag{8}$$

Due to the Pauli statistics of local excitations resonant pairs interact with each other as spins *1/2*. Indeed, two flip-flop states can be described by a single spin *1/2* with two possible states $\sigma_{ij}^z = \pm 1/2$ corresponding to the states $|\psi_\pm>$ in Eq. (4), respectively. The longitudinal interaction $U_{ik}S_i^z S_k^z + U_{il}S_i^z S_l^z + U_{jk}S_j^z S_k^z + U_{jl}S_j^z S_l^z$ of two pairs $(i,j)$ and $(k,l)$ Eq. (2) stimulates two simultaneous flip-flop transitions (Fig. 2) in both pairs $V_{ij}^{lk}\sigma_{ij}^x \sigma_{kl}^x$ characterized by the amplitude

$$V_{ij}^{lk} = (U_{ik} - U_{il} - U_{jk} + U_{jl})\sin(\varphi_{ij})\sin(\varphi_{kl})/2. \qquad (9)$$

The longitudinal interaction also leads to $\sigma_{ij}^z \sigma_{kl}^z, \sigma_{ij}^z \sigma_{kl}^x$ terms which does not affect resonant probability because they act similarly to a random field $\Delta$.[11] The transverse interaction does not couple different pairs with each other so we can ignore it. Since we consider resonant pairs both *sin(φ)* factors in Eq. (9) are of order of *1* and we can skip them in our estimates, assuming that the "flip-flop" interaction of pairs (Fig. 2) is equivalent to the longitudinal interaction Eq. (1).

Can the many-body interaction Eq. (9) of resonant pairs stimulate the delocalization of energy? To answer this question we perform the scaling analysis of interacting pairs at different pair energies *E*. This scaling starts with the characteristic interaction energy $E \sim \xi U_0 n^{a/d}$ with subsequent reduction of *E* down to *0*. If the interaction at energy *E* is weak compared to the characteristic pair energy *E* then one can exclude the pairs with energy ~*E* as immobile and proceed to lower energy pairs. The excluded high energy pairs and other excitations contribute to static disordering of low energy pairs, which does not affect the qualitative estimate for a number of resonances.[11] Consider the coupling of resonant pairs of the certain energy $E < \xi U_0 n^{a/d}$. The density of such pairs is given by Eq. (8) and their coupling strength at the distance *R* is $U_0/R^a$ Eq. (9). When the coupling strength of two pairs exceeds their typical energy *E* they are in resonance to each other (strong coupling). All neighbours of the given pair separated from it less than $R_E = (U_0/E)^{1/a}$ defined by the condition $U_0/R_E^a \sim E$ are in resonance with it. The number of resonant neighbours (cf. Eq. (6))

$$\chi_E = n_p(E) R_E^d = \frac{U_0 n^{a/d}}{W} \left( \frac{\xi U_0 n^{a/d}}{E} \right)^{\frac{2d-a}{a}} \qquad (10)$$

Eq. (10) shows that when the interaction $U(r) \sim r^{-a}$ decreases with the distance slower than $1/R^{2d}$ (*a<2d*) the coupling strength of resonant pairs approaches infinity with decreasing their energy $E$. When $\chi_p(E) \gg 1$ disordering is weak and each pair is coupled in resonant manner with $\chi_p(E)$ neighbours by strong flip-flop interactions. Therefore at $a<2d$ the inevitable delocalization of energy must take place.[3] The percolation condition $\chi_p(E_*) = \chi_c \sim 1$ determines the energy bandwidth $E_*$ of delocalized excitations. The numerical parameter $\chi_c$ is less than unity and is of order of $10^{-1}$ Eq. (6). Then we get $E_* = U_0 n^{a/d} \left( \frac{\xi U_0 n^{a/d}}{W \chi_c^a} \right)^{\frac{1}{2d-a}} \left( \frac{U_0 n^{a/d}}{E} \right)^{\frac{2d-a}{d}}$. Since this energy defines both the typical excitation transition rate in resonant pairs $1/t_* \sim E_*/\hbar$ and the energy fluctuation induced by this transition Eq. (7) the associated rate $1/t_*$ defines the system decoherence rate. Indeed, the irreversible phase fluctuation of any spin during the time $t_*$ is of order of $1$ ($\delta\phi \sim \delta E t_*/\hbar \sim E_*(\hbar/E_*)/\hbar \sim 1$) which means that decoherence takes place during the time $t_*$.[5] Thus one can estimate the decoherence rate of interacting spins as

$$\frac{1}{\tau_2} = \frac{U_0 n^{a/d}}{\hbar} \left( \frac{\xi U_0 n^{a/d}}{W \chi_c^a} \right)^{\frac{1}{2d-a}} \left( \frac{U_0 n^{a/d}}{E} \right)^{\frac{2d-a}{d}}. \qquad (11)$$

Let us briefly resume our findings. We show that the decoherence always takes place in the infinite system of interacting spins (qubits) containing finite fraction of excited spins if the interaction decreases with the distance slower then $1/r^{2d}$. In quantum computers the destructive effect of interaction stimulated decoherence can be avoided at times less than the the decoherence time $\tau_2$ Eq. (11). Alternatively one can use the small system of the size less than the critical size $L_*$ defined by the delocalization energy $E_*$ as $U_0/L_*^a = E_*$. The long-range interaction will be not significant in such a small system so it becomes decoherence-free with respect to the interaction of qubits.

It is interesting that dipolar ($a=3$) and quadrupole ($a=5$) interactions in *3-d* systems and a dipolar interaction ($a=3$) in *2-d* system inevitably lead to the energy delocalization at finite temperature ($a<2d$). Consequently interaction of localized electrons in *3-d* and *2-d* doped semiconductors should lead to an energy delocalization and consequently a charge conduction at finite temperature despite of Anderson localization of electrons at *T=0*.

The irreversible energy transport inevitably leads to the relaxation of all excitations including non-resonant ones. This problem requires special study. The investigation of relaxation mechanisms following the scenario of Refs. [5, 6] is in progress.

This work is supported by the Louisiana Board of Regents, Contract No. LEQSF 2005-08-RDA-29. Author acknowledges for useful discussions, criticism and suggestions Leonid Maksimov, Yuri Kagan, Il'ya Polishchuk, Clare Yu, Boris Shklovskii, Philip Stamp, Igor Tupitzyn and Yuri Galperin. I also wish to specially acknowledge Boris Shklovskii for his hospitality in the University of Minnesota where the remarkable part of this work has been done during the hurricane disaster in New Orleans (Fall 2005).

1. Anderson, P. W., Absence of Diffusion in Certain Random Lattices, *Phys. Rev.* **109**, 1492-1505 (1958).
2. Levitov, L. S. Critical Hamiltonians with long range hopping. *Annal. der Physik* **8**, 697-706 (1999); Aharonov, D., Kitaev, A., Preskill, J. Fault-Tolerant Quantum

Computation with Long-Range Correlated Noise. *Phys. Rev. Lett.* **96**, 050504/1-4 (2006).

3. Delocalization in low-dimensions $d=1, 2$ always requires a special consideration because of the full localization of all states in the corresponding single particle problem. Since the delocalization parameter Eq. (10) can be made arbitrarily high for sufficiently small energy, coherent backscattering for such excitations will take long time. During that time energy fluctuations caused by other excitations will breakdown the coherence similarly to the case of an electron localization breakdown in low-dimension induced by electronic interaction at finite temperature.[7]

4. Galperin, Y.M., Gurevich V.L., Parshin, D. A., Nonlinear high-frequency hopping conductivity of semiconductors and spectral diffusion, Zh. Exp. and Teor. Fiz. (USSR), **94**, 364-382 (1988).

5. See Burin, A.L., Kagan, Yu., Maksimov, L.A., Polishchuk, I.Y. Dephasing Rate in Dielctric Glasses at Ultralow Temperatures, Phys. Rev. Lett. **80**, 2945 (1998) and the review Burin, A. L., Natelson, D., Osheroff, D. D., Kagan, Y. Interactions between tunneling defects in amorphous solids. in "*Tunneling systems in amorphous and crystalline solids*", eds. P. Esquinazi, Springer-Verlag, Berlin, Heidelberg, New York, 1998 and references therein.

6. Prokofev, N.V., Stamp, P.C.E. Low-Temperature Quantum Relaxation in a System of Magnetic Nanomolecules. *Phys. Rev. Lett.* **80**, 5794-5797 (2005).

7. Yu, C. C., Leggett, A. J. Low temperature properties of amorphous materials: through a glass darkly. *Comments Cond. Matt. Phys.* **14**, 231-251 (1988).

8. Nielsen, A., Chuang, I.L. Quantum computation and quantum information, Cambridge, U. K. (New York: Cambridge University Press, 2000); Johnson, G. A shortcut through time: the path to the quantum computer (New York: Alfred A. Knopf, 2003).

9. Ohtsuki, T., Slevin, K., Kawarabayashi, T. Review of recent progress on numerical studies of the Anderson localization. *Ann. Phys.* (Leipzig) **8**, 655-664 (1999).

10. Grussbach, H., Schreiber, M., Determination of the mobility edge in the Anderson model of localization in three dimensions by multifractal analysis, *Phys. Rev. B* **51**, 663 (1995).

11. In this letter we ignore off-resonant interactions. This can be justified as following. Omitted longitudinal terms can be included into external disordering and they cannot affect the probability of resonance because the characteristic disordering energy exceeds interaction Eq. (4). Off-diagonal transverse terms like $\sigma^z \sigma^x$ can increase the number of possible final states after flip-flop transition stimulated by the transverse interaction, because other spins can also overturn. Then the flip-flop transition amplitude $V$ decouples into the sequence of amplitudes $V_1, V_2, \ldots V_n$, such as $V^2 = \sum_{i=1}^{n} V_i^2$ for different channels all leading to states with different energies $E_1, \ldots E_n$. This effect always increases the probability of resonance because for $n$ resonances it is given by $\frac{1}{W}\sum_{i=1}^{n}|V_i| > \frac{|V|}{W}$ similarly to previously studied delocalization of the Floquet states, Burin, A.L., Kagan, Yu., Polishchuk, I. Ya. Energy Transport Induced by an External Alternating Field in Strongly Disordered Media. *Phys. Rev. Lett.* **86**, 5616-5619 (2001). In our case of interest it can be shown applying perturbation theory methods (see Burin, A.L., Kontor, K.N., Maksimov, L.A. Localization and Delocalization in the Paramagnetic Phase of the Transverse Ising Model. *Theor. Math. Phys.* **85**, 1223 (1990) and later work Basko, D. M., Aleiner, I. L., Altshuler, B. L. Metal-insulator transition in a weakly interacting many-electron system with localized single-particle states. *Annals of Physics* **321**, 1126-1205 (2006); Gornyi, I. V., Mirlin, A. D., Polyakov, D. G. Interacting electrons in disordered wires. Anderson localization and low-T transport. *Phys. Rev. Lett.* **95**, Art. No. 206603 (2005), where similar methods were applied to electrons in doped semiconductors in the model of the short range interaction) that off-resonant terms does not affect the delocalization transition estimate.

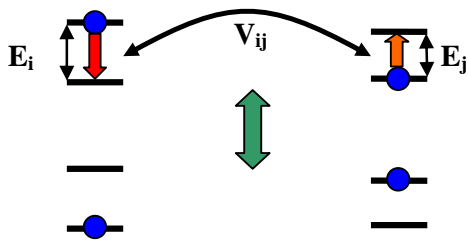

**Fig. 1. Flip-flop transition of a single particle excitation under the condition of resonance $E_i-E_j < V_{ij}$**

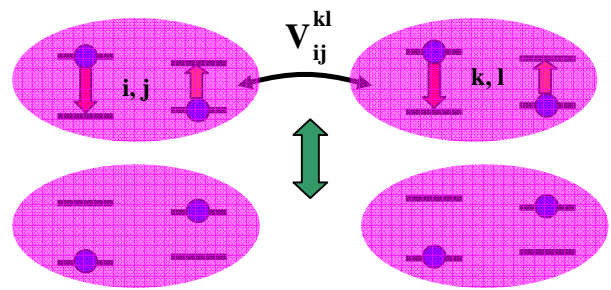

**Fig. 2. Flip-flop transition of two pairs.**